\documentclass[prb,twocolumn,longbibliography]{revtex4-1}
\pdfoutput=1

\usepackage{amsmath}
\usepackage{amsfonts}
\usepackage{graphicx}
\usepackage{color}

\newcommand{\real}{\operatorname{Re}}
\newcommand{\imag}{\operatorname{Im}}

\begin{document}

\title{Comment on ``The negative flow of probability''}

\author{Arseni Goussev}

\affiliation{School of Mathematics and Physics, University of Portsmouth, Portsmouth PO1 3HF, United Kingdom}

\date{\today}

\begin{abstract}
	The left-to-right motion of a free quantum Gaussian wave packet can be accompanied by the right-to-left flow of the probability density, the effect recently studied by Villanueva [Am.~J.~Phys.~{\bf 88}, 325 (2020)]. Using the Wigner representation of the wave packet, we analyze the effect in phase space, and demonstrate that its physical origin is rooted in classical mechanics.
\end{abstract}

\maketitle

In a recent paper \cite{Vil20negative}, Villanueva has explored a seemingly paradoxical effect associated with the motion of a free quantum particle. The particle state is represented by a Gaussian wave packet:
\begin{equation}
	\psi(x,t) = \exp \left( \frac{i}{\hbar} \alpha_t (x - x_t)^2 + \frac{i}{\hbar} p_0 (x - x_t) + \frac{i}{\hbar} \gamma_t \right) \,,
\label{psi}
\end{equation}
where $x$ and $t$ are the position and time variables, respectively. Here,
\begin{equation}
	x_t = x_0 + \frac{p_0 t}{m} \,,
\label{x_t}
\end{equation}
is the wave packet center at time $t$, with $x_0$ and $p_0$ being the initial mean position and momentum, respectively; $m$ is the particle mass. The complex-valued function $\alpha_t$ is defined as 
\begin{equation}
	\frac{1}{\alpha_t} = \frac{1}{\alpha_0} +  \frac{2 t}{m} \,,
\label{alpha_t}
\end{equation}
and controls the wave packet spread,
\begin{equation*}
	(\Delta x)_t \equiv \sqrt{ \langle x^2 \rangle_t - \langle x \rangle_t^2} =  \frac{1}{2} \sqrt{\frac{\hbar}{\imag \alpha_t}} \,,
\end{equation*}
along with position-momentum correlations. Its initial value, $\alpha_0$, must satisfy $\imag \alpha_0 > 0$ for the wave packet to be normalizable. Finally, the complex-valued function $\gamma_t$ encapsulates both the normalization constant and global phase; the imaginary part of $\gamma_t$ is related to that of $\alpha_t$ via
\begin{equation}
	\exp \left( -\frac{2 \imag \gamma_t}{\hbar} \right) = \sqrt{\frac{2 \imag \alpha_t}{\pi \hbar}} \,.
\label{gamma}
\end{equation}
The wave function $\psi(x,t)$ satisfies the free particle Schr\"odinger equation:
\begin{equation}
	i \hbar \frac{\partial \psi}{\partial t} = -\frac{\hbar^2}{2 m} \frac{\partial^2 \psi}{\partial x^2} \,.
\label{TDSE}
\end{equation}

The effect addressed by Villanueva can be summarized as follows. Let $q$ be some fixed point on the $x$-axis, and consider the scenario in which
\begin{equation*}
	q - x_0 \gg (\Delta x)_0 \qquad \text{and} \qquad p_0 > 0 \,.
\end{equation*}
In other words, the initial wave packet is localized (almost entirely) on the left of $q$ and is moving to the right. Naively, one might think that, as the wave packet approaches the point $q$ from the left (that is, as long as $x_t < q$), the probability of finding the particle in the region $x > q$, namely
\begin{equation}
	\Pi(t) = \int_q^{+\infty} dx \, |\psi(x,t)|^2 \,,
\label{Pi_psi}
\end{equation}
grows monotonically with time, i.e. one might expect that $\frac{d}{d t} \Pi(t) > 0$ for $0 < t < t_{\text{cl}} \equiv \frac{m (q - x_0)}{p_0}$. However, as Villanueva demonstrates, this naive intuition can be false: If the value of the parameter $\alpha_0$ is such that
\begin{equation}
	\real \alpha_t < -\frac{p_0}{2 (q - x_t)} \,,
\label{condition}
\end{equation}
at an instant $t < t_{\text{cl}}$, then
the probability $\Pi(t)$ appears to be decreasing,
\begin{equation*}
	\frac{d}{d t} \Pi(t) < 0 \,.
\end{equation*}

Here we point out that, if considered in phase space, the above effect has a simple intuitive explanation. In fact, the effect is rooted in classical mechanics: The same negative flow of probability takes place in an ensemble of free classical particles with an appropriate Gaussian distribution of positions and momenta. We also present a phase-space-based derivation of condition~\eqref{condition}.

Let us regard the motion of a free particle, described in the Schr\"odinger picture by the wave function $\psi(x,t)$, as the time evolution of the Wigner phase-space (quasiprobability) density \cite{Sny80Use, Cas08Wigner}
\begin{equation*}
	W(x,p,t) = \frac{1}{\pi \hbar} \int_{-\infty}^{+\infty} dy \, e^{-i 2 p y / \hbar} \psi(x + y, t) \psi^*(x - y, t) \,.
\end{equation*}
 The reader is referred to review articles \cite{Sny80Use, Cas08Wigner} for a discussion of many fascinating properties of the Wigner density. Here, we only state the following two properties, particularly relevant to the present discussion. First, the time evolution of $W(x,p,t)$ in free space is governed by
\begin{equation}
	W(x,p,t) = W(x - p t / m, p, 0) \,.
\label{W_vs_t}
\end{equation}
This equation is the phase-space representation to the free-particle Schr\"odinger equation~\eqref{TDSE}. Second, in terms of the Wigner function, the probability of finding the particle in the region $x > q$ reads
\begin{equation}
	\Pi(t) = \int_q^{+\infty} dx \int_{-\infty}^{+\infty} dp \, W(x,p,t) \,.
\label{Pi_W}
\end{equation}
This is the phase-space representation of  Eq.~\eqref{Pi_psi}.

Substituting $\psi(x,t)$, given by Eq.~\eqref{psi}, into Eq.~\eqref{W_vs_t}, performing the integration, and using identity~\eqref{gamma}, we obtain
\begin{equation}
	W(x,p,t) = \frac{1}{\pi \hbar} \exp \left( -\frac{2 \widetilde{x}^2 \imag \alpha_t}{\hbar} - \frac{\left( \widetilde{p} - 2 \widetilde{x} \real \alpha_t \right)^2}{2 \hbar \imag \alpha_t} \right) \,,
\label{W_Gaussian}
\end{equation}
where
\begin{equation*}
	\widetilde{x} = x - x_t \qquad \text{and} \qquad \widetilde{p} = p - p_0
\end{equation*}
are, respectively, the particle position and momentum measured relative to their mean values. The Wigner function~\eqref{W_Gaussian} is positive at all times, $W(x,p,t) > 0$.

Now comes an important (and well known) argument. Imagine an ensemble of $N$ free noninteracting {\it classical} particles of mass $m$. We are interested in the limit of large $N$. Let $N W_{\text{cl}}(x,p,t)$ be the number density of the particles at the phase-space point $(x,p)$ at time $t$. So, $W_{\text{cl}}(x,p,t)$ is the probability density for a given particle to be found at $(x,p)$ at time $t$. Since the momentum of a free particle is conserved, the time evolution of $W_{\text{cl}}$ is given by
\begin{equation}
	W_{\text{cl}}(x,p,t) = W_{\text{cl}}(x - p t / m, p, 0) \,.
\label{W_cl_vs_t}
\end{equation}
The number of particles in the region $x > q$ at time $t$ equals $N \Pi_{\text{cl}}(t)$, where
\begin{equation}
	\Pi_{\text{cl}}(t) = \int_q^{+\infty} dx \int_{-\infty}^{+\infty} dp \, W_{\text{cl}}(x,p,t)
\label{Pi_W_cl}
\end{equation}
represents the probability for a given particle to satisfy $x > q$ at time $t$. Observe that Eqs.~\eqref{W_vs_t} and \eqref{Pi_W} for the Wigner function $W$ of a quantum particle are identical to Eqs.~\eqref{W_cl_vs_t} and \eqref{Pi_W_cl} for the classical phase-space probability density $W_{\text{cl}}$, respectively \footnote{The equivalence between Eq.~\eqref{W_vs_t} and Eq.~\eqref{W_cl_vs_t} is a particular case of the following fact: If the system potential is of the form $V(x) = a x^2 + b x + c$, where $a$, $b$, and $c$ are some constants, then the quantum evolution equation for the Wigner function is identical to the classical Liouville equation for the phase-space probability density. The free particle case addressed in the present paper corresponds to $a = b = 0$.}. It then immediately follows that
\begin{equation}
	\Pi(t) = \Pi_{\text{cl}}(t) \,,
\end{equation}
provided that $W(x,p,0) = W_{\text{cl}}(x,p,0)$. In general \cite{Sny80Use, Cas08Wigner}, the set of all possible Wigner functions $W(x, p, 0)$ is not the same as the set of all possible classical probability densities $W_{\text{cl}}(x, p, 0)$. For example, $W(x,p,0)$ can have negative values, whereas $W_{\text{cl}}(x,p,0)$ is nonnegative by construction; on the other hand, the value of $W_{\text{cl}}(x,p,0)$ can in principle be arbitrarily large, whereas $|W(x,p,0)| \le 1 / \pi \hbar$ for all $x$ and $p$. However, the Wigner function $W$ representing a {\it Gaussian} wave packet is everywhere positive, Eq.~\eqref{W_Gaussian}, and therefore can be regarded as a valid classical probability density $W_{\text{cl}}$. This guaranties that the behavior of $\Pi(t)$ for a Gaussian quantum state, described by an initial Wigner function $W(x,p,0)$, is identical to the behavior of $\Pi_{\text{cl}}(t)$ for an ensemble of free classical particles, initially distributed in accordance with the phase-space density $W_{\text{cl}}(x,p,0) = W(x,p,0)$. In particular, this means that {\it the negative flow of probability addressed by Villanueva \cite{Vil20negative} is essentially a classical-mechanical effect.}

We now present a phase-space interpretation of the effect. For the discussion below, it is important to introduce dimensionless versions of the position, momentum, and time variables. This requires defining a natural length scale $L$. We achieve this by noticing that, according to Eq.~\eqref{alpha_t}, the quantity $\imag \left( 1 / \alpha_t \right) = -(\imag \alpha_t) / |\alpha_t|^2$ does not depend on time, i.e.~$\imag \left( 1 / \alpha_t \right) = \imag \left( 1 / \alpha_0 \right)$. Thus, $L$ can be defined as
\begin{equation}
	L^2 \equiv \frac{\hbar}{2} \frac{\imag \alpha_t}{|\alpha_t|^2} = \frac{\hbar}{2} \frac{\imag \alpha_0}{|\alpha_0|^2} \,.
\label{dimless-L}
\end{equation}
Subsequently, we introduce dimensionless positions
\begin{equation}
	\xi = \frac{x}{L} \,, \qquad \xi_0 = \frac{x_0}{L} \,, \qquad \delta = \frac{q}{L} \,,
\label{dimless-xi}
\end{equation}
dimensionless momenta
\begin{equation}
	\eta = \frac{p}{\hbar / L} \,, \qquad \eta_0 = \frac{p_0}{\hbar / L} \,,
\label{dimless-eta}
\end{equation}
dimensionless time
\begin{equation*}
	\tau = \frac{t}{m L^2 / \hbar} \,,
\end{equation*}
and dimensionless (quasi)probability density
\begin{equation*}
	\Omega(\xi, \eta, \tau) = \frac{W(x, p, t)}{1 / \hbar} \,.
\end{equation*}
A straightforward calculation yields the following dimensionless version of the Wigner function \eqref{W_Gaussian}:
\begin{equation}
	\Omega(\xi, \eta, \tau) = \frac{1}{\pi} e^{-\left( \widetilde{\xi} - \epsilon_{\tau} \widetilde{\eta} \right)^2 - \widetilde{\eta}^2} \,,
\label{Omega}
\end{equation}
where
\begin{equation}
	\epsilon_{\tau} = \left. \frac{\real \alpha_t}{\imag \alpha_t} \, \right|_{t = \frac{m L^2}{\hbar} \tau} \,,
\label{dimless-eps}
\end{equation}
and
\begin{equation}
	\widetilde{\xi} = \xi - \xi_\tau \qquad \text{and} \qquad \widetilde{\eta} = \eta - \eta_0 \,.
\label{xi_and_eta_tilde}
\end{equation}
The time dependence of the dimensionless Wigner function is specified by
\begin{equation}
	\xi_{\tau} = \xi_0 + \eta_0 \tau \,,
\label{xi_tau}
\end{equation}
cf.~Eq.~\eqref{x_t}, and
\begin{equation}
	\epsilon_{\tau} = \epsilon_0 + \tau \,,
\label{epsilon_tau}
\end{equation}
cf.~Eq.~\eqref{alpha_t}, where $\epsilon_0 = \real \alpha_0 / \imag \alpha_0$. It is easy to check that the dimensionless Wigner function satisfies the free particle evolution equation
\begin{equation}
	\Omega(\xi, \eta, \tau) = \Omega(\xi - \eta \tau, \eta, 0) \,,
\label{Omega_vs_tau}
\end{equation}
cf.~Eqs.~\eqref{W_vs_t} and \eqref{W_cl_vs_t}, and that the probability of finding the particle in the region $\xi > \delta$ (or $x > q$) at time $\tau$ (or $t$) is given by
\begin{equation*}
	\Pi(\tau) = \int_\delta^{+\infty} d\xi \int_{-\infty}^{+\infty} d\eta \, \Omega(\xi, \eta, \tau) \,,
\end{equation*}
cf.~Eqs.~\eqref{Pi_W} and \eqref{Pi_W_cl}.

\begin{figure}[h]
	\centering
	\includegraphics[width=0.4\textwidth]{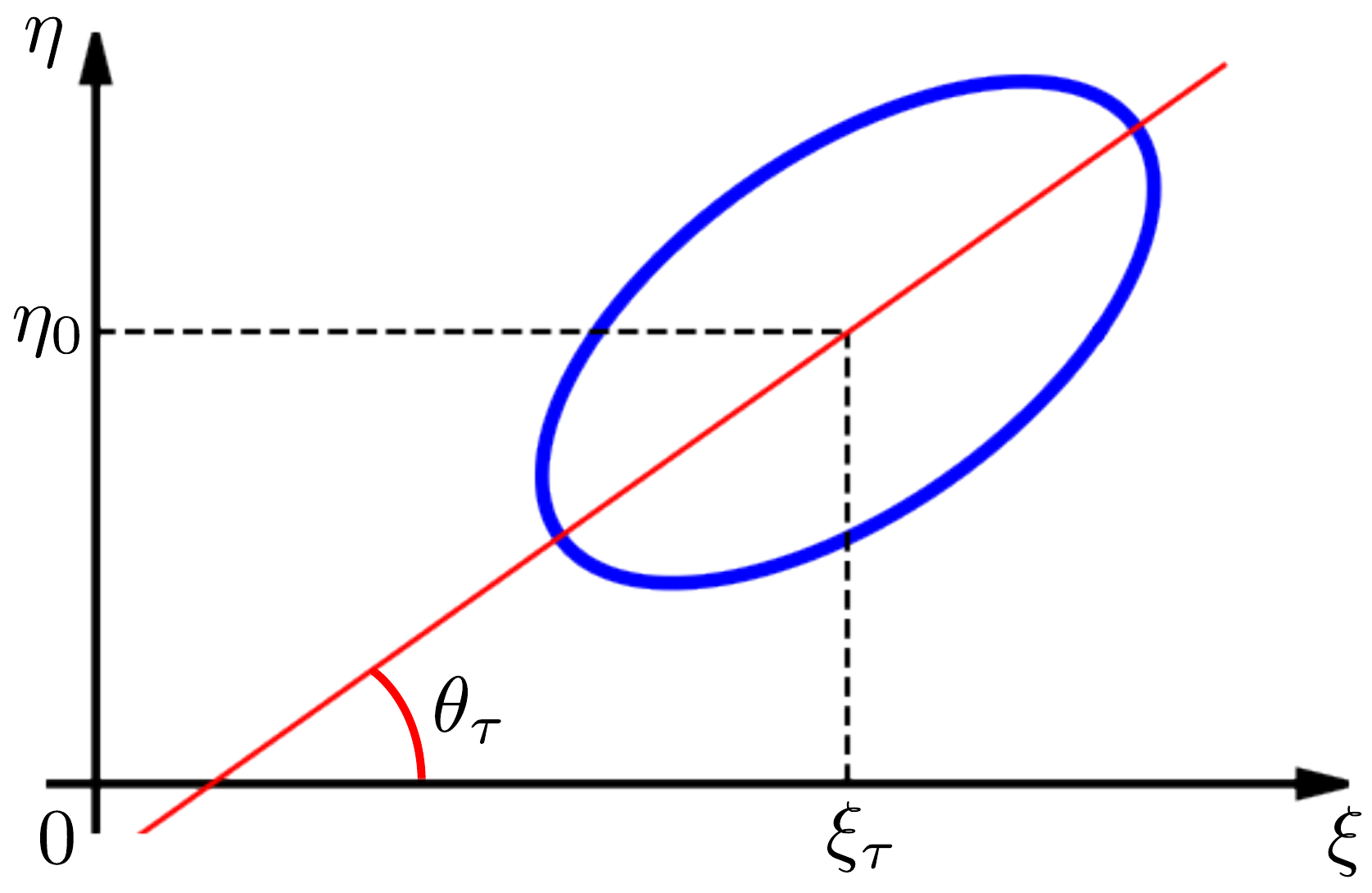}
	\caption{An elliptical curve in phase space on which the Wigner function $\Omega(\xi, \eta, \tau)$ has a constant value. The ellipse becomes a circle when $\epsilon_{\tau} = 0$; this corresponds to a minimal uncertainty state.}
	\label{fig1}
\end{figure}
\begin{figure}[h]
	\centering
	\includegraphics[width=0.45\textwidth]{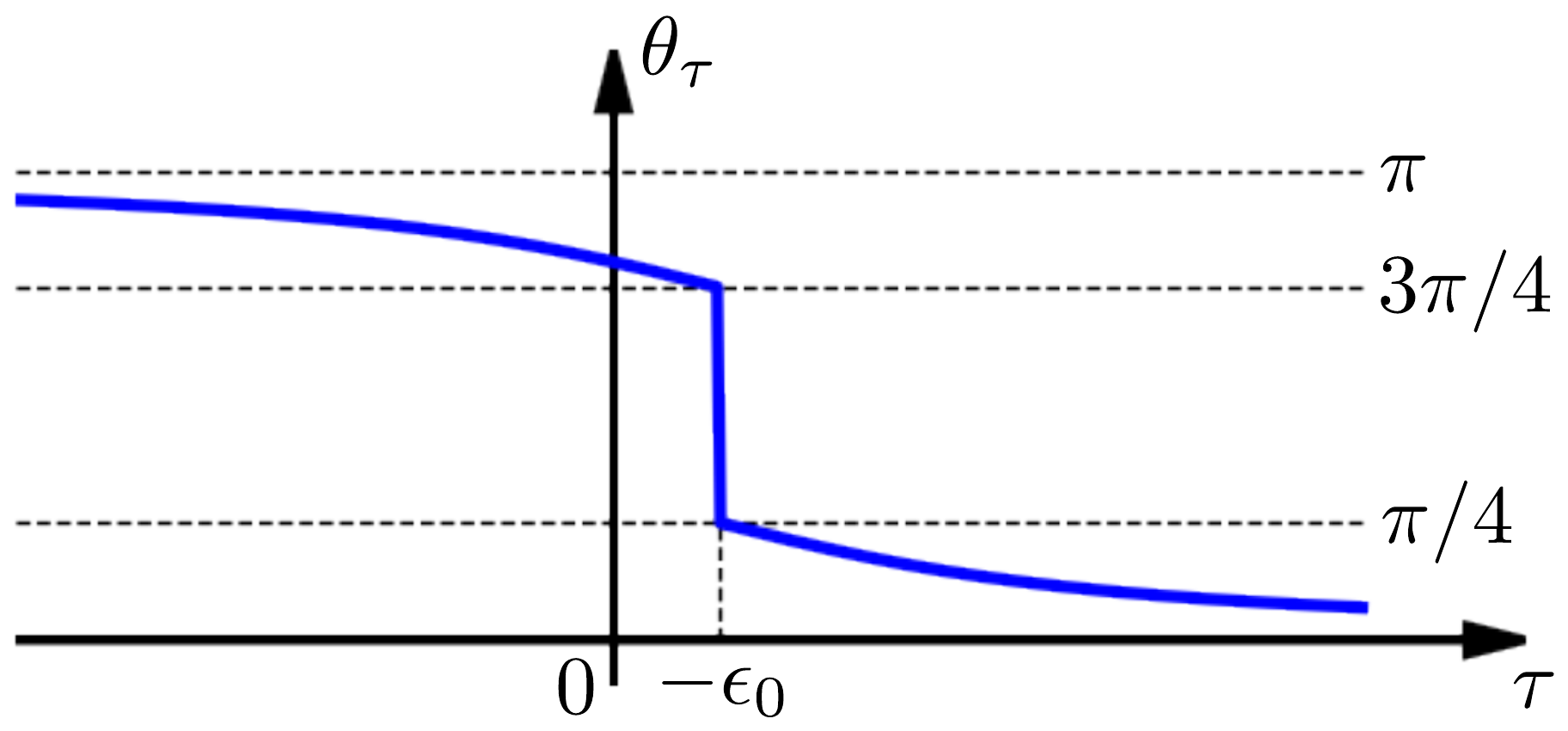}
	\caption{Angle $\theta_{\tau}$, illustrated in Fig.~\ref{fig1}, as a function of time $\tau$. At $\tau = -\epsilon_0$, the particle is in a minimal uncertainty state, for which $\theta_{\tau}$ is not defined.}
	\label{fig2}
\end{figure}
Equations~\eqref{Omega}, \eqref{xi_and_eta_tilde}, \eqref{xi_tau}, and \eqref{epsilon_tau} provide the complete description of the motion of a free Gaussian wave packet. At time $\tau$, the wave packet is parametrized by three dimensionless real numbers: $\xi_{\tau}$, $\eta_0$, and $\epsilon_{\tau}$. Figure~\ref{fig1} shows a phase-space curve of a constant value of $\Omega(\xi, \eta, \tau)$. The curve is an ellipse centered at $(\xi_{\tau}, \eta_0)$. It is easy to show (see Appendix~\ref{appendix1}) that the angle between the major axis of the ellipse and the $\xi$-axis is given by
\begin{equation}
	\theta_{\tau} = \left\{
	\begin{array}{ll}
		\displaystyle \frac{1}{2} \arctan \frac{2}{\epsilon_{\tau}} &\text{if} \quad \epsilon_{\tau} > 0 \\[0.4cm]
		\displaystyle \pi + \frac{1}{2} \arctan \frac{2}{\epsilon_{\tau}} \quad &\text{if} \quad \epsilon_{\tau} < 0
	\end{array}
	\right. \,.
\label{theta}
\end{equation}
The angle $\theta_{\tau}$ decreases monotonically from $\pi$ to $0$ as time $\tau$ increases from $-\infty$ to $+ \infty$. This is shown in Fig.~\ref{fig2}. According to Eq.~\eqref{epsilon_tau}, $\epsilon_{\tau} = 0$ when time $\tau = -\epsilon_0$. At this instant, the Wigner function representing the wave packet reads
\begin{equation*}
	\Omega(\xi, \eta, -\epsilon_0) = \frac{1}{\pi} e^{-\tilde{\xi}^2 - \tilde{\eta}^2} \,.
\end{equation*}
This is a minimal uncertainty state. The phase-space contour lines representing minimal uncertainty states are circles, and so the angle $\theta_{\tau}$ is not defined at $\tau = -\epsilon_0$. In Fig.~\ref{fig2}, the value of $\epsilon_0$ is chosen to be negative.

\begin{figure}[h]
	\centering
	\includegraphics[width=0.39\textwidth]{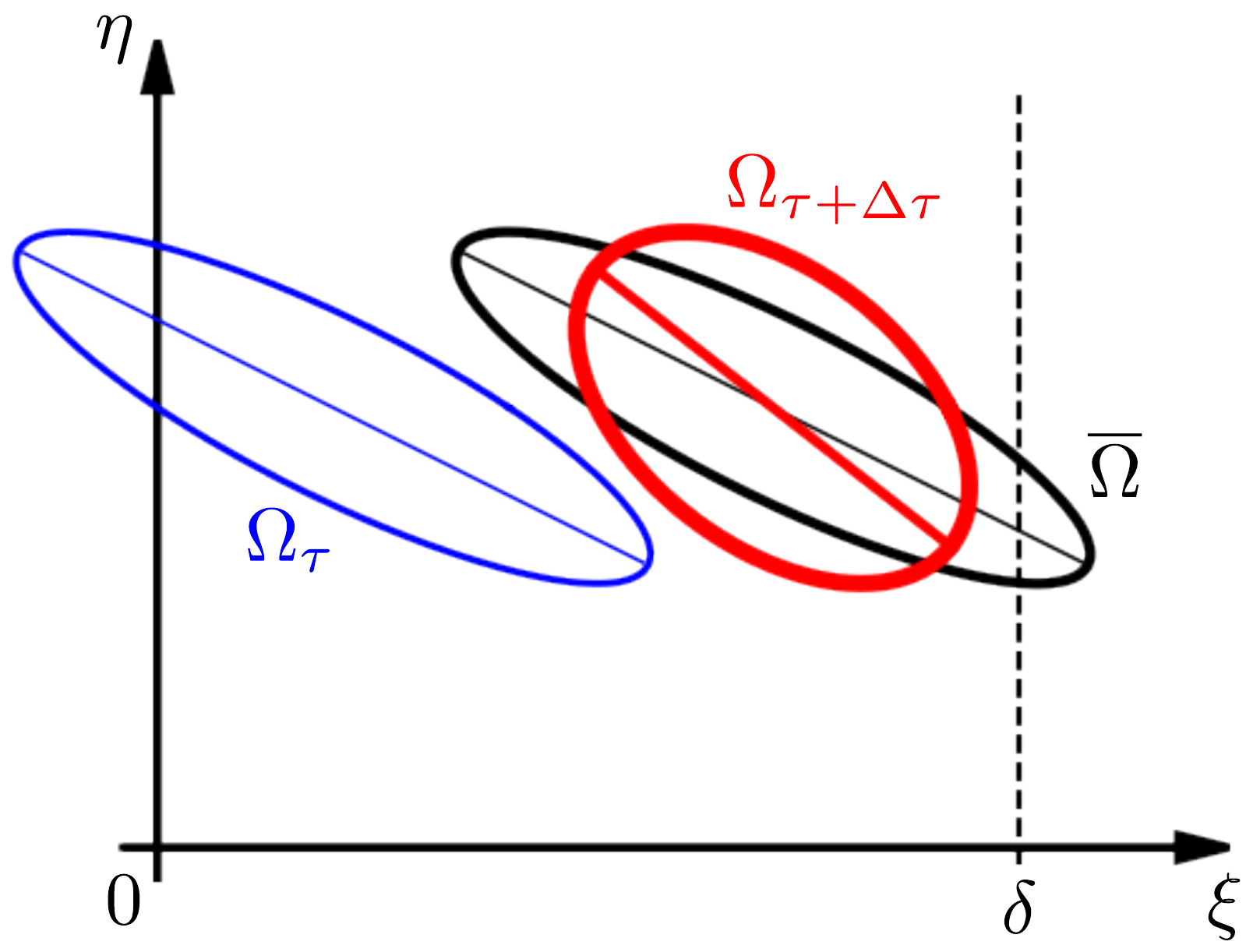}
	\caption{The time evolution of the Wigner function from $\Omega_{\tau} \equiv \Omega(\xi, \eta, \tau)$ to $\Omega_{\tau + \Delta \tau} \equiv \Omega(\xi, \eta, \tau + \Delta \tau)$ as a sequence two consecutive transformations: $\Omega_{\tau} \to \overline{\Omega}$ and $\overline{\Omega} \to \Omega_{\tau + \Delta \tau}$. See the text for details. Each Wigner function is represented by an elliptical contour line, along with the corresponding major axis. The left boundary of the spatial region $\xi > \delta$ is shown with a dashed line.}
	\label{fig3}
\end{figure}
The time evolution of the Wigner function during the time interval between $\tau$ and $\tau + \Delta\tau$, i.e.
\begin{equation*}
	\Omega(\xi, \eta, \tau) \quad \to \quad \Omega(\xi, \eta, \tau + \Delta \tau) = \Omega(\xi - \eta \Delta \tau, \eta, \tau) \,,
\end{equation*}
can be viewed as the result of two consecutive transformations, both illustrated in Fig.~\ref{fig3}. The first transformation is a rigid shift of the Wigner distribution along the $\xi$-axis by $\eta_0 \Delta \tau$:
\begin{equation}
	\Omega(\xi, \eta, \tau) \quad \to \quad \overline{\Omega}(\xi, \eta) = \Omega(\xi - \eta_0 \Delta \tau, \eta, \tau) \,.
\label{transform_1}
\end{equation}
The second transformation is a simple shear leaving the distribution center unchanged:
\begin{equation}
	\overline{\Omega}(\xi, \eta) \quad \to \quad \Omega(\xi, \eta, \tau + \Delta \tau) = \overline{\Omega} \big( \xi - (\eta-\eta_0) \Delta \tau, \eta \big) \,.
\label{transform_2}
\end{equation}
The change in the Wigner density $\Delta \Omega$ during the time interval between $\tau$ and $\tau + \Delta \tau$ is given by
\begin{align*}
	\Delta \Omega
	&\equiv \Omega(\xi, \eta, \tau + \Delta \tau) - \Omega(\xi, \eta, \tau) \\
	&= (\Delta \Omega)_{\text{shift}} + (\Delta \Omega)_{\text{shear}} \,,
\end{align*}
where
\begin{equation}
	(\Delta \Omega)_{\text{shift}} = \overline{\Omega}(\xi, \eta) - \Omega(\xi, \eta, \tau)
\label{Delta_Omega_shift}
\end{equation}
is the change due to the shift transformation, and
\begin{equation}
	(\Delta \Omega)_{\text{shear}} = \Omega(\xi, \eta, \tau + \Delta \tau) - \overline{\Omega}(\xi, \eta)
\label{Delta_Omega_shear}
\end{equation}
is the change due to the shear transformation. Consequently, the corresponding change in the probability of finding the particle in the region $\xi > \delta$ is
\begin{align*}
	\Delta \Pi
	&\equiv \Pi(\tau + \Delta \tau) - \Pi(\tau) \\
	&= \int_\delta^{+\infty} d\xi \int_{-\infty}^{+\infty} d\eta \, \Delta \Omega \\
	&= (\Delta \Pi)_{\text{shift}} + (\Delta \Pi)_{\text{shear}} \,,
\end{align*}
where
\begin{equation}
	(\Delta \Pi)_{\text{shift}} = \int_\delta^{+\infty} d\xi \int_{-\infty}^{+\infty} d\eta \, (\Delta \Omega)_{\text{shift}}
\label{Delta_Pi_shift_def}
\end{equation}
and
\begin{equation}
	(\Delta \Pi)_{\text{shear}} = \int_\delta^{+\infty} d\xi \int_{-\infty}^{+\infty} d\eta \, (\Delta \Omega)_{\text{shear}} \,.
\label{Delta_Pi_shear_def}
\end{equation}
It is straightforward to show (see Appendix~\ref{appendix2}) that, in the limit of small $\Delta \tau$,
\begin{equation}
	(\Delta \Pi)_{\text{shift}} = \eta_0 \Delta \tau \int_{-\infty}^{+\infty} d\eta \, \Omega(\delta, \eta, \tau)
\label{Delta_Pi_shift}
\end{equation}
and
\begin{equation}
	(\Delta \Pi)_{\text{shear}} = \Delta \tau \int_{-\infty}^{+\infty} d\eta \, (\eta - \eta_0) \Omega(\delta, \eta, \tau) \,.
\label{Delta_Pi_shear}
\end{equation}

In the case of the Wigner function $\Omega$ given by Eq.~\eqref{Omega}, the probability change due to the shift transformation is always positive:
\begin{equation*}
	(\Delta \Pi)_{\text{shift}} > 0 \,.
\end{equation*}
This follows directly from the fact that $\Omega(\xi, \eta, \tau) > 0$. However, the probability change due to the shear transformation, $(\Delta \Pi)_{\text{shear}}$, can be negative for some values of the wave packet parameters. Figure~\ref{fig3} provides an example of the situation in which $(\Delta \Pi)_{\text{shear}} < 0$; it is clear that $\epsilon_{\tau} < 0$ (or, equivalently, $\theta_{\tau}  > 3 \pi/4$) is a necessary condition for $(\Delta \Pi)_{\text{shear}}$ to be negative. The negative flow of the net probability, $\Delta \Pi < 0$, occurs if and only if
\begin{equation}
	(\Delta \Pi)_{\text{shear}} < -(\Delta \Pi)_{\text{shift}} \,.
\label{negative_flow_condition}
\end{equation}
This condition is the phase-space equivalent of Eq.~\eqref{condition}.

In order to explicitly recover Eq.~\eqref{condition} from Eq.~\eqref{negative_flow_condition}, we substitute Eq.~\eqref{Omega} into Eqs.~\eqref{Delta_Pi_shift} and \eqref{Delta_Pi_shear}, and evaluate the corresponding momentum integrals. This yields (see Appendix~\ref{appendix2})
\begin{equation}
	(\Delta \Pi)_{\text{shift}} = \frac{\eta_0 \Delta \tau}{\sqrt{\pi \left( 1 + \epsilon_{\tau}^2 \right)}} \exp \left[ -\frac{(\delta - \xi_{\tau})^2}{1 + \epsilon_{\tau}^2} \right]
\label{Delta_Pi_shift_Gaussian}
\end{equation}
and
\begin{equation}
	(\Delta \Pi)_{\text{shear}} = \frac{\epsilon_{\tau} (\delta - \xi_{\tau}) \Delta \tau}{\sqrt{\pi \left( 1 + \epsilon_{\tau}^2 \right)^3}} \exp \left[ -\frac{(\delta - \xi_{\tau})^2}{1 + \epsilon_{\tau}^2} \right] \,.
\label{Delta_Pi_shear_Gaussian}
\end{equation}
Substituting these expressions into Eq.~\eqref{negative_flow_condition}, we obtain
\begin{equation}
	\frac{\epsilon_{\tau}}{1 + \epsilon_{\tau}^2} < -\frac{\eta_0}{\delta - \xi_{\tau}} \,.
\label{condition_dimensionless}
\end{equation}
Using the rescaling transformations~\eqref{dimless-L}, \eqref{dimless-xi}, \eqref{dimless-eta}, and \eqref{dimless-eps}, it can be easily verified that Eq.~\eqref{condition_dimensionless} is the dimensionless version of the negative probability flow condition~\eqref{condition}. The geometrical meaning of this condition is discussed in Appendix~\ref{appendix3}.

In summary, we have shown that the effect of the negative probability flow, studied by Villanueva \cite{Vil20negative}, has an intuitive explanation when considered in phase space. The effect is classical-mechanical in nature, and occurs not only for a free quantum particle with a Gaussian wave function, but also for an ensemble of free classical particles with a Gaussian distribution of positions and momenta.

%%%%%%%%%%%%%%%%%%%%%%%%%%%%%%%
\appendix

\section{Derivation of Eq.~\eqref{theta}}
\label{appendix1}

There are many ways to derive Eq.~\eqref{theta}. Here we adopt a direct one. It follows from Eq.~\eqref{Omega} that phase-space points $(\xi, \eta)$ corresponding to the same value of $\Omega(\xi, \eta, \tau)$ lie on the ellipse
\begin{equation*}
	\left( \widetilde{\xi} - \epsilon_{\tau} \widetilde{\eta} \right)^2 + \widetilde{\eta}^{2} = C \,,
\end{equation*}
where $C > 0$ is a constant. In polar coordinates,
\begin{equation*}
	\widetilde{\xi} = r \cos \theta \,, \qquad \widetilde{\eta} = r \sin \theta \,,
\end{equation*}
the ellipse equation takes the form
\begin{equation*}
	r^2 = \frac{C}{1 - \epsilon_{\tau} \sin 2 \theta + \epsilon_{\tau}^2 \sin^2 \theta} \,.
\end{equation*}
The angle $\theta = \theta_{\tau}$ is the one that maximizes $r$, and therefore satisfies the equation
\begin{equation*}
	\frac{d r^2}{d \theta} = 0 \,,
\end{equation*}
which is equivalent to
\begin{equation*}
	2 \epsilon_{\tau} \cos 2 \theta - \epsilon_{\tau}^2 \sin 2 \theta = 0 \,.
\end{equation*}
For $\epsilon_{\tau} \not= 0$, this reduces to
\begin{equation*}
	\tan 2 \theta = \frac{2}{\epsilon_{\tau}} \,.
\end{equation*}
On the interval $0 < \theta < \pi$, this equation has two solutions: (i) the solution given by Eq.~\eqref{theta}, and (ii) the one representing the orthogonal direction, i.e.
\begin{equation*}
	\theta_{\tau}^{\perp} = \frac{\pi}{2} + \frac{1}{2} \arctan \frac{2}{\epsilon_{\tau}} \,.
\end{equation*}
A straightforward evaluation of $\frac{d^2 r}{d \theta^2}$ shows that it is the solution given by Eq.~\eqref{theta} that maximizes $r$. 

\section{Derivation of Eqs.~\eqref{Delta_Pi_shift}, \eqref{Delta_Pi_shear}, \eqref{Delta_Pi_shift_Gaussian}, \eqref{Delta_Pi_shear_Gaussian}}
\label{appendix2}

We first derive Eqs.~\eqref{Delta_Pi_shift} and \eqref{Delta_Pi_shear}. From Eqs.~\eqref{Delta_Omega_shift} and \eqref{transform_1}, we have
\begin{align*}
	(\Delta \Omega)_{\text{shift}}
	&= \Omega(\xi - \eta_0 \Delta \tau, \eta, \tau) - \Omega(\xi, \eta, \tau) \\
	&= -\eta_0 \Delta \tau \frac{\partial \Omega(\xi, \eta, \tau)}{\partial \xi} + O \left( \Delta \tau^2 \right) \,.
\end{align*}
Substituting this into Eq.~\eqref{Delta_Pi_shift_def}, and only keeping the leading order term in $\Delta \tau$, we obtain
\begin{align*}
	(\Delta \Pi)_{\text{shift}}
	&= -\eta_0 \Delta \tau \int_{-\infty}^{+\infty} d\eta \int_{\delta}^{+\infty} d\xi \, \frac{\partial \Omega(\xi, \eta, \tau)}{\partial \xi} \\
	&= \eta_0 \Delta \tau \int_{-\infty}^{+\infty} d\eta \, \Omega(\delta, \eta, \tau) \,,
\end{align*}
where we have taken into account the fact that $\Omega(\xi, \eta, \tau) \to 0$ as $\xi \to \infty$. Similarly, from Eqs.~\eqref{Delta_Omega_shear}, \eqref{transform_2}, and \eqref{transform_1}, we have
\begin{align*}
	(\Delta \Omega)_{\text{shear}}
	&= \overline{\Omega}\big( \xi - (\eta - \eta_0) \Delta, \eta \big) - \overline{\Omega}(\xi, \eta) \\
	&= -(\eta - \eta_0) \Delta \tau \frac{\partial \overline{\Omega}(\xi, \eta)}{\partial \xi} + O \left( \Delta \tau^2 \right) \\
	&= -(\eta - \eta_0) \Delta \tau \frac{\partial}{\partial \xi} \Omega(\xi - \eta_0 \Delta \tau, \eta, \tau) + O \left( \Delta \tau^2 \right) \\
	&= -(\eta - \eta_0) \Delta \tau \frac{\partial \Omega(\xi, \eta, \tau)}{\partial \xi} + O \left( \Delta \tau^2 \right) \,.
\end{align*}
Substituting this into Eq.~\eqref{Delta_Pi_shear_def}, and only keeping the leading order term in $\Delta \tau$, we obtain
\begin{align*}
	(\Delta \Pi)_{\text{shear}}
	&= -\Delta \tau \int_{-\infty}^{+\infty} d\eta \, (\eta - \eta_0) \int_{\delta}^{+\infty} d\xi \, \frac{\partial \Omega(\xi, \eta, \tau)}{\partial \xi} \\
	&= \Delta \tau \int_{-\infty}^{+\infty} d\eta \, (\eta - \eta_0) \Omega(\delta, \eta, \tau) \,.
\end{align*}

Now we derive Eq.~\eqref{Delta_Pi_shift_Gaussian} and \eqref{Delta_Pi_shear_Gaussian}. According to Eq.~\eqref{Omega}, $\Omega$ can be written as
\begin{equation*}
	\Omega(\xi, \eta, \tau) = \frac{1}{\pi} e^{-a \eta^2 + 2 b \eta - c}
\end{equation*}
with
\begin{equation*}
	a = 1 + \epsilon_{\tau}^2 \,,
\end{equation*}
\begin{equation*}
	b = \left( 1 + \epsilon_{\tau}^2 \right) \eta_0 + \epsilon_{\tau} (\delta - \xi_{\tau}) \,,
\end{equation*}
\begin{equation*}
	c = (\delta - \xi_{\tau} + \epsilon_{\tau} \eta_0)^2 + \eta_0^2 \,.
\end{equation*}
Hence,
\begin{equation*}
	\int_{-\infty}^{+\infty} d\eta \, \Omega(\delta, \eta, \tau) = \frac{1}{\pi} \int_{-\infty}^{+\infty} d\eta \, e^{-a \eta^2 + 2 b \eta - c} = \frac{e^{b^2 / a - c}}{\sqrt{\pi a}} \,.
\end{equation*}
Since
\begin{equation*}
	\frac{b^2}{a} - c = -\frac{(\delta - \xi_{\tau})^2}{1 + \epsilon_{\tau}^2} \,,
\end{equation*}
we get
\begin{equation*}
	\int_{-\infty}^{+\infty} d\eta \, \Omega(\delta, \eta, \tau) = \frac{1}{\sqrt{\pi \left( 1 + \epsilon_{\tau}^2 \right)}} \exp \left[ -\frac{(\delta - \xi_{\tau})^2}{1 + \epsilon_{\tau}^2} \right] \,,
\end{equation*}
which, in turn, leads to Eq.~\eqref{Delta_Pi_shift_Gaussian}. Then,
\begin{align*}
	\int_{-\infty}^{+\infty} d\eta \, \eta \Omega(\delta, \eta, \tau) &= \frac{1}{\pi} \int_{-\infty}^{+\infty} d\eta \, \eta e^{-a \eta^2 + 2 b \eta - c} \\
	&= \frac{b}{a} \frac{e^{b^2 / a - c}}{\sqrt{\pi a}} \,,
\end{align*}
and so
\begin{align*}
	\int_{-\infty}^{+\infty} d\eta \, (\eta - \eta_0) &\Omega(\delta, \eta, \tau)
	= (b - a \eta_0) \frac{e^{b^2 / a - c}}{\sqrt{\pi a^3}} \\
	&= \frac{\epsilon_{\tau} (\delta - \xi_{\tau})}{\sqrt{\pi \left( 1 + \epsilon_{\tau}^2 \right)^3}} \exp \left[ -\frac{(\delta - \xi_{\tau})^2}{1 + \epsilon_{\tau}^2} \right] \,.
\end{align*}
This yields Eq.~\eqref{Delta_Pi_shear_Gaussian}.

\section{Geometrical meaning of Eq.~\eqref{condition_dimensionless}}
\label{appendix3}

\begin{figure}[h]
	\centering
	\includegraphics[width=0.37\textwidth]{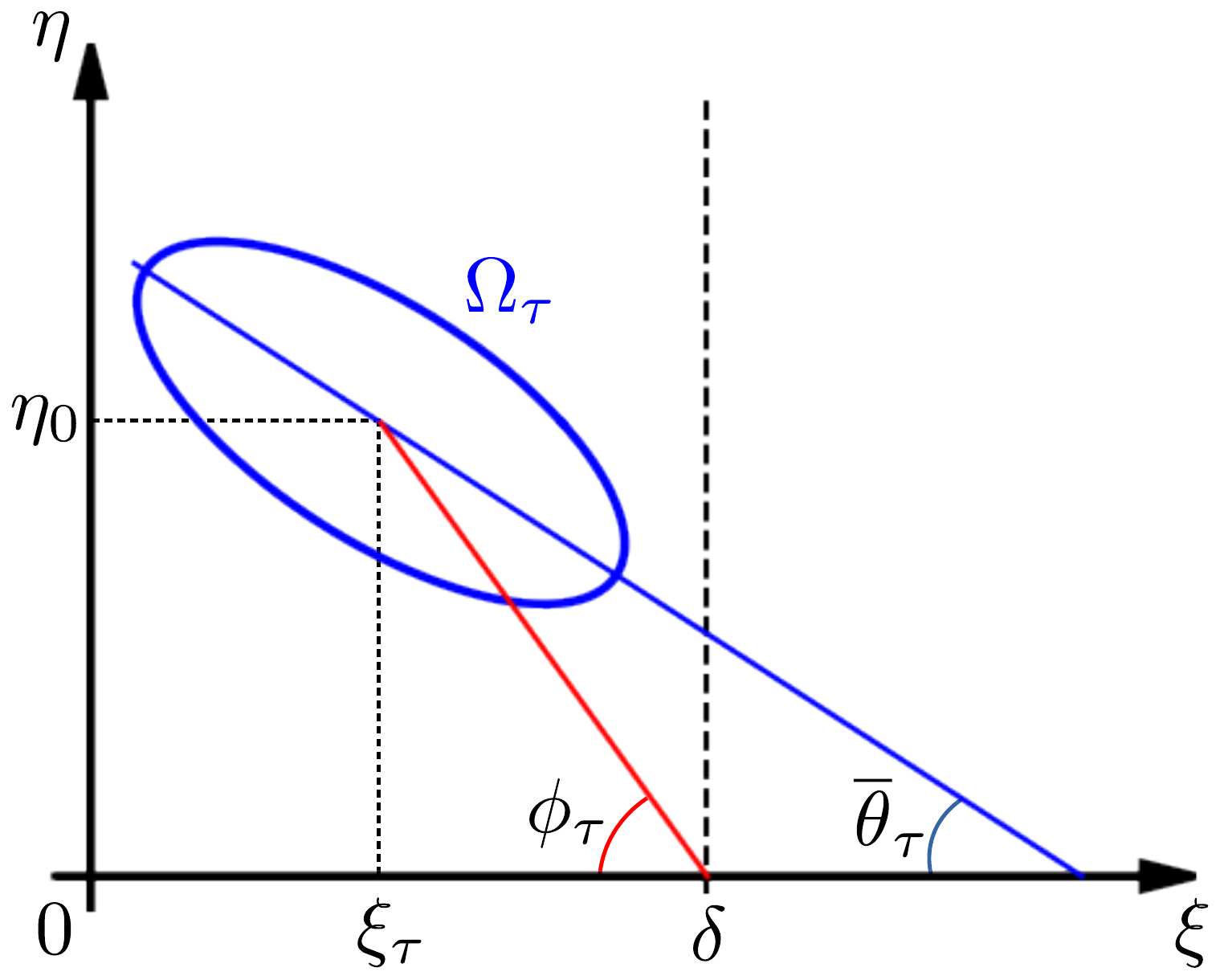}
	\caption{Angles $\overline{\theta}_{\tau} = \pi - \theta_{\tau}$ and $\phi_{\tau}$ for the Wigner function $\Omega_{\tau} \equiv \Omega(\xi, \eta, \tau)$.}
	\label{fig4}
\end{figure}

Condition~\eqref{condition_dimensionless} determines the phase-space position and orientation of the Wigner function $\Omega(\xi,\eta,\tau)$ accompanied by negative probability flow at the point $\xi = \delta$. The geometrical meaning of this condition becomes clear if we rewrite Eq.~\eqref{condition_dimensionless} in terms of angles $\overline{\theta}_{\tau} = \pi - \theta_{\tau}$ and $\phi_{\tau}$, defined in Fig.~\ref{fig4}. We have
\begin{equation*}
	\frac{\eta_0}{\delta - \xi_{\tau}} = \tan \phi_{\tau} > 0
\end{equation*}
and, in view of Eq.~\eqref{theta},
\begin{equation*}
	\epsilon_{\tau} = \frac{2}{\tan 2 \theta_{\tau}} = -\frac{2}{\tan 2 \overline{\theta}_{\tau}} \,.
\end{equation*}
Substituting these two expressions into Eq.~\eqref{condition_dimensionless}, we find that $\overline{\theta}_{\tau}$ must lie inside the interval $(0, \pi/4)$ and satisfy the inequality
\begin{equation}
	\frac{\tan 2 \overline{\theta}_{\tau}}{2} + \frac{2}{\tan 2 \overline{\theta}_{\tau}} < \frac{1}{\tan \phi_{\tau}} \,.
\label{condition_angles}
\end{equation}
This inequality is equivalent to Eq.~\eqref{condition_dimensionless}.

\begin{figure}[h]
	\centering
	\includegraphics[width=0.45\textwidth]{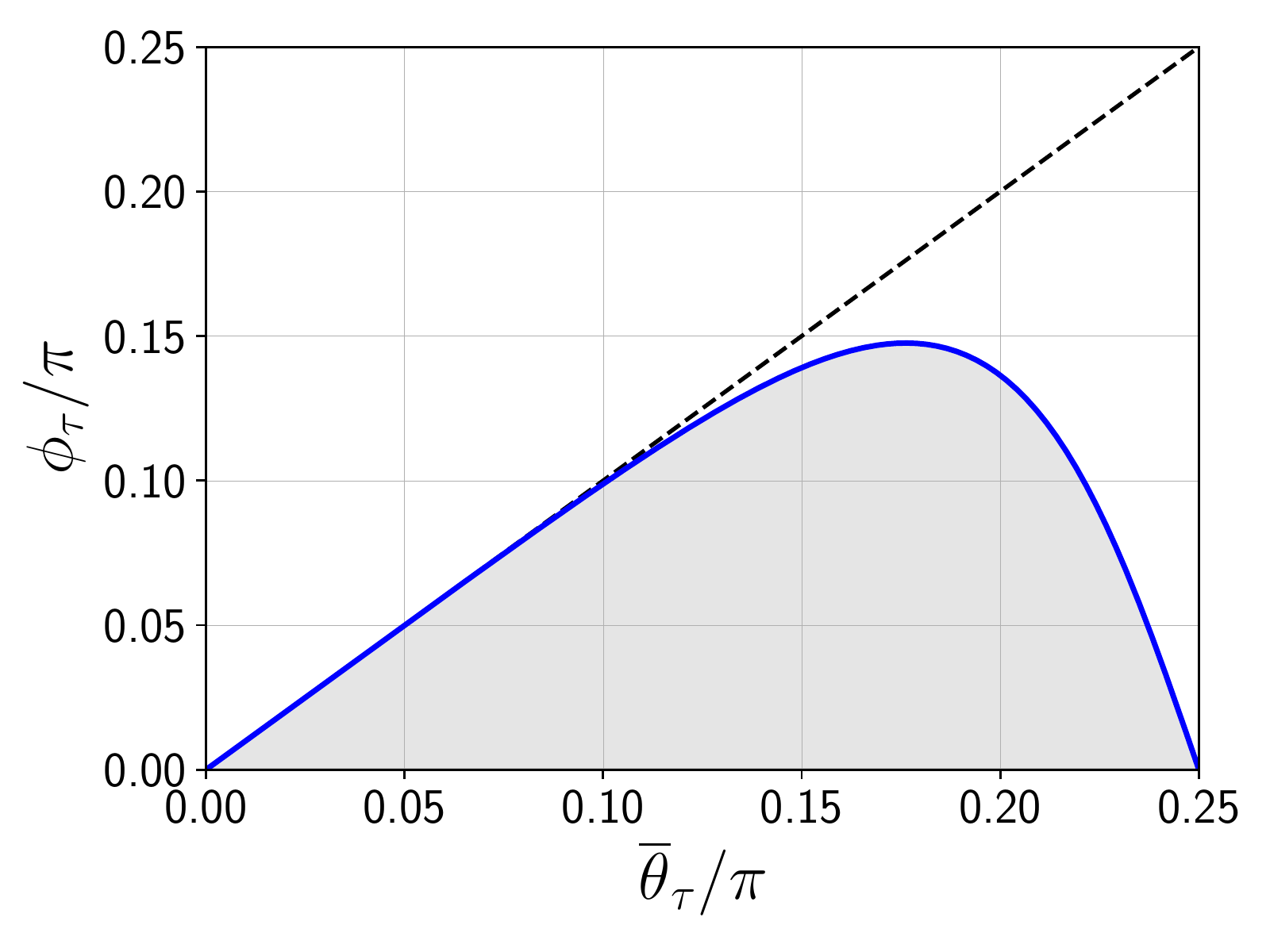}
	\caption{The shaded area shows the angles $\overline{\theta}_{\tau}$ and $\phi_{\tau}$ for which condition~\eqref{condition_angles} is fulfilled. The dashed line corresponds to $\phi_{\tau} = \overline{\theta}_{\tau}$.}
	\label{fig5}
\end{figure}

Figure~\ref{fig5} shows the set of angle pairs $(\overline{\theta}_{\tau}, \phi_{\tau})$ fulfilling condition~\eqref{condition_angles}. The condition becomes especially simple in the limit of small $\overline{\theta}_{\tau}$: If $\overline{\theta}_{\tau} \ll 1$, then negative probability flow at $\xi = \delta$ occurs for $\phi_{\tau} < \overline{\theta}_{\tau}$.

%\bibliography{../this_paper}

%

\end{document}